\documentclass[]{interact}

\usepackage{epstopdf}
\usepackage[caption=false]{subfig}

\usepackage[numbers,sort&compress]{natbib}
\bibpunct[, ]{[}{]}{,}{n}{,}{,}
\renewcommand\bibfont{\fontsize{10}{12}\selectfont}
\makeatletter
\def\NAT@def@citea{\def\@citea{\NAT@separator}}
\makeatother

\theoremstyle{plain}

\theoremstyle{definition}

\theoremstyle{remark}

\begin{document}


\title{{\em A Revolution in Science}: the Eclipse Expeditions of 1919}

\author{
\name{Peter Coles\thanks{CONTACT Peter Coles. Email: Peter.Coles@mu.ie}}
\affil{Department of Theoretical Physics, Maynooth University, Maynooth, County Kildare, Ireland}
}

\maketitle

\begin{abstract}
  The first direct experimental test of Einstein's theory of
  general relativity involved  a pair of expeditions to measure the bending of light at a total solar eclipse that
  took place one hundred years ago, on 29th May 1919.
So famous is this experiment, and so dramatic was the impact on
Einstein himself, that history tends not to recognize the
controversy that surrounded the results at the time. In this article
I discuss the experiment in its  scientific and historical background
context and explain why it was, and is, such an important episode in the development
of modern physics.
 \end{abstract}


\section{Introduction}

In the past century, the discoveries in the field of
physical science have unfolded at a remarkable rate.
Physicists have unravelled the structure
of matter on the tiniest accessible scales, breaking up atomic
nuclei into elementary particles and studying the forces that
cause these particles to interact.  Over the same period
astronomers discovered have established that the Universe is expanding,
and cosmologists are now trying to understand the very instant of creation at the Big Bang
that started off this expansion. These daring adventures of the
mind which a hundred years ago would have seemed fanciful are
based on solid foundations of experiment, observation and
theory.

The era of modern physics in which we find ourselves
began with Galileo and Newton. But the early years of the twentieth century saw a
dramatic acceleration in this progress. In particular, in 1919 an experiment was
performed that was intended to test Einstein's general theory of
relativity. The science involved in this experiment provides an interesting way to introduce
some of the key concepts of general relativity as well as some aspects
of statistical data analysis. It therefore has considerable pedagogical
value to this day. Moreover, the results caused an unprecedented
media sensation and turned Albert
Einstein into a household name almost overnight. As well as providing an
illustration of of some of the key concepts underlying Einstein's theory, the story of this experiment
and its aftermath  also reveals interesting insights into the
relationship between science and wider society, which I shall touch on at the end of this article.

\section{Universal Gravitation}

\subsection{Newton's Laws}
Gravity is one of the four fundamental forces of nature. It represents
the universal tendency of all matter to attract all other matter.
This universality sets it apart from, for example, the forces
between electrically-charged bodies, because electrical charges
can be of two different kinds, positive or negative. While
electrical forces can lead either to attraction (between unlike
charges) or repulsion (between like charges), gravity is always
attractive.

In many ways, the force of gravity is extremely weak. Most
material bodies are held together by electrical forces between
atoms which are many orders of magnitude stronger than the
gravitational forces between them. But, despite its weakness,
gravity is the driving force in astronomical situations because
astronomical bodies, with very few exceptions, always contain
exactly the same amount of positive and negative charge and
therefore never exert forces of an electrical nature on each
other.

One of the first great achievements of theoretical physics was
Isaac Newton's theory of universal gravitation, which unified what
had seemed to be many disparate physical phenomena. Newton's
theory of mechanics is encoded in three simple laws:
\begin{enumerate}
\item Every body continues in a state of rest or uniform motion
in a straight line unless it is compelled to change that state by
forces impressed upon it.
\item Rate of change of momentum is
proportional to the impressed force, and is in the direction in
which this force acts.
\item To every action, there is always
opposed an equal reaction.
\end{enumerate}
These three laws of motion are general, applying just as
accurately to the behaviour of balls on a billiard table as to the
motion of the heavenly bodies. All that Newton needed to do was to
figure out how to describe the force of gravity. Newton realised
that a body orbiting in a circle, like the Moon going around the
Earth, is experiencing a force in the direction of the centre of
motion (just as a weight tied to the end of a piece of string does
when it is twirled around one's head). Gravity could cause this
motion in the same way as it could cause apples to fall to Earth
from trees. In both these situations, the force has to be towards
the centre of the Earth. Newton realised that the correct form of
mathematical equation was an inverse-square law:
\begin{equation}
F=G\frac{M_{\rm A}M_{\rm B}}{r^2}.
\end{equation}
In other words the attractive force $F$ between any two bodies of
masses $M_A$ and $M_B$ depends on the product of the masses of the
bodies and upon the square of the distance $r$ between them. The
quantity $G$ is a fundamental constant, called Newton's constant.
Combined with Newton's Second Law relating the force $F$ on a body
to its acceleration $a$:
\begin{equation}
F=M~\,a,
\end{equation}
this allows one to calculate the changes in motion and position of
bodies interacting under gravity.

It was a triumph of Newton's theory that the inverse-square law of
universal gravitation could explain the laws of planetary motion
obtained by Johannes Kepler more than a century earlier. So
spectacular was this success that the idea of a Universe guided by
Newton's laws of motion was to dominate scientific thinking for
more than two centuries. Until, in fact, the arrival on the scene
of an obscure patent clerk by the name of Albert Einstein.

\subsection{The Einstein Revolution}
A detailed account of the life of Albert Einstein can be found in
Pais (1992). He was born in Ulm (Germany) on 14 March 1879, but
his family soon moved to Munich, where he spent his school years.
The young Einstein was not a particularly good student, and in
1894 he dropped out of school entirely when his family moved to
Italy. After failing the entrance examination once, he was
eventually admitted to the Swiss Institute of Technology in Zurich
in 1896. Although he did fairly well as a student in Zurich, he
was unable to get a job in any Swiss university, as he was held to
be extremely lazy. He left academia to work in the Patent Office
at Bern in 1902. This gave him a good wage and, since the tasks
given to a junior patent clerk were not exactly onerous, it also
gave him plenty of spare time to think about physics.

Einstein's special theory of relativity stands as
one of the greatest intellectual achievements in the history of
human thought. It is made even more remarkable by the fact that
Einstein was still working as a patent clerk at the time, and was
only doing physics as a kind of hobby. What's more, he also
published seminal works that year on the photoelectric effect and
on Brownian motion. But the reason why the special theory of
relativity stands head-and-shoulders above his own work of this
time, and that of his colleagues in the world of mainstream
physics, is that Einstein managed to break away completely from
the concept of time as an absolute property that marches on at the
same rate for everyone and everything. This idea is built into the
Newtonian picture of the world, and most of us regard it as being
so obviously true that it does not bear discussion.

The idea of relativity did not originate with Einstein. The
principle of it had been articulated by Galileo nearly three
centuries earlier. Galileo claimed that only relative motion
matters, so there could be no such thing as absolute motion. He
argued that if you were travelling in a boat at constant speed on
a smooth lake, then there would be no experiment that you could do
in a sealed cabin on the boat that would indicate to you that you
were moving at all. Of course, not much was known about physics in
Galileo's time, so the kinds of experiment he could envisage were
rather limited.

Einstein's version of the principle of relativity simply turned it
into the statement that all laws of nature have to be exactly the
same for all observers in relative motion. In particular, Einstein
decided that this principle must apply to the theory of
electromagnetism, constructed by James Clerk Maxwell, which
describes amongst other things the forces between charged bodies
mentioned above. One of the consequences of Maxwell's theory is
that the speed of light (in vacuum) appears as a universal
constant $c$. Taking the principle of relativity seriously means
that all observers have to measure the same value of c, whatever
their state of motion. This seems straightforward enough, but the
consequences are nothing short of revolutionary.

\subsection{Thought Experiments}
Einstein decided to ask himself specific questions about what
would be observed in particular kinds of experiments involving the
exchange of light signals. He worked a great deal with {\em
gedanken} (thought) experiments of this kind. For example, imagine
there is a flash bulb in the centre of a railway carriage moving
along a track. At each end of the carriage there is a clock, so
that when the flash illuminates it we can see the time. If the
flash goes off, then the light signal reaches both ends of the
carriage simultaneously, from the point of view of passengers
sitting in the carriage. The same time is seen on each clock.

Now picture what happens from the point of view of an observer at
rest who is watching the train from the track. The light flash
travels with the same speed in our reference frame as it did for
the passengers. But the passengers at the back of the carriage are
moving into the signal, while those at the front are moving away
from it. This observer therefore sees the clock at the back of the
train light up before the clock at the front does. But when the
clock at the front does light up, it reads the same time as the
clock at the back did! This observer has to conclude that
something is wrong with the clocks on the train. This example
demonstrates that the concept of simultaneity is relative. The
arrivals of the two light flashes are simultaneous in the frame of
the carriage, but occur at different times in the frame of the
track. Other examples of strange relativistic phenomena include
time dilation (moving clocks appear to run slow) and length
contraction (moving rulers appear shorter). These are all
consequences of the assumption that the speed of light must be the
same as measured by all observers. Of course, the examples given
above are a little unrealistic. In order to show noticeable
effects, the velocities concerned must be a sizeable fraction of
c. Such speeds are unlikely to be reached in railway carriages.
Nevertheless, experiments have been done that show that time
dilation effects are real. The decay rate of radioactive particles
is much slower when they are moving at high velocities because
their internal clock runs slowly. Special relativity also spawned
the most famous equation in all physics
\begin{equation}
 E=mc^{2},
 \end{equation}
expressing the equivalence between matter and energy.

Remarkable though the special theory undoubtedly is, it is
seriously incomplete. It deals only with bodies moving with
constant velocity with respect to each other. Even Chapter One of
the laws of Nature, written by Newton, had been built around the
causes and consequences of velocities that change with time.
Newton's second law is about the rate of change of momentum of an
object, which in layman's terms is its acceleration. Special
relativity is restricted to so-called inertial motions, i.e. the
motions of particles that are not acted upon by any external
forces. This means that special relativity cannot describe
accelerated motion of any kind and, in particular, cannot describe
motion under the influence of gravity.

\subsection{The Equivalence Principle}
Einstein had a number of deep insights in how to incorporate
gravitation into relativity theory. For a start, consider Newton's
theory of gravity embodied by Equation (1). The force exerted on
body B by body A depends on $M_{\rm A}$ and $M_{\rm B}$. In this
case mass has the role of a kind of gravitational charge,
determining the strength of the pull. Consequently, this
manifestation of mass is called the gravitational mass; the force
produced depends on the {\em active} gravitational mass (in this case
$M_{\rm A}$ and the force felt by B depends on its {\em passive}
gravitational mass $M_{\rm B}$. But we then have to use Newton's
second law (2) to work out the acceleration. The acceleration of B
depends then on $M_{\rm B}$, but mass plays a different role in
this expression. In Equation (2) we have the {\em inertial mass}
of the particle which represents its reluctance to being
accelerated.  But Newton's third law of motion also states that if
body A exerts a force on body B then body B exerts a force on body
A which is equal and opposite. This means that m must also be the
active gravitational mass (if you like, the gravitationa {\em charge})
produced by the particle. In Newton's theory, all three of these
masses - the inertial mass, the active and passive gravitational
masses - are equivalent. But there seems to be no reason, on the
face of it, why this should be the case. Couldn't they be
different?

Einstein decided that this equivalence must be the consequence of
a deeper principle called the principle of equivalence. In his own
words, this means that 'all local, freely-falling laboratories are
equivalent for the performance of all physical experiments'. What
this means is essentially that one can do away with gravity as a
separate force of nature and regard it instead as a consequence of
moving between accelerated frames of reference.

 To see how this is possible, imagine a lift
equipped with a physics laboratory. If the lift is at rest on the
ground floor, experiments will reveal the presence of gravity to
the occupants. For example, if we attach a weight on a spring to
the ceiling of the lift, the weight will extend the spring
downwards. Next, imagine that we take the lift to the top of a
building and let it fall freely. Inside the freely-falling lift
there is no perceptible gravity. The spring does not extend, as
the weight is always falling at the same rate as the rest of the
lift, even though the lift's speed might be changing. This is what
would happen if we took the lift out into space, far away from the
gravitational field of the Earth. The absence of gravity therefore
looks very much like the state of free-fall in response to a
gravitational force. Moreover, imagine that our lift was actually
in space (and out of gravity's reach), but there was a rocket
attached to it. Firing the rocket would make the lift accelerate.
There is no up or down in free space, but let us assume that the
rocket is attached so that the lift would accelerate in the
opposite direction from before, i.e. in the direction of the
ceiling.

\begin{figure}
\centering
\resizebox*{10cm}{!}{\includegraphics{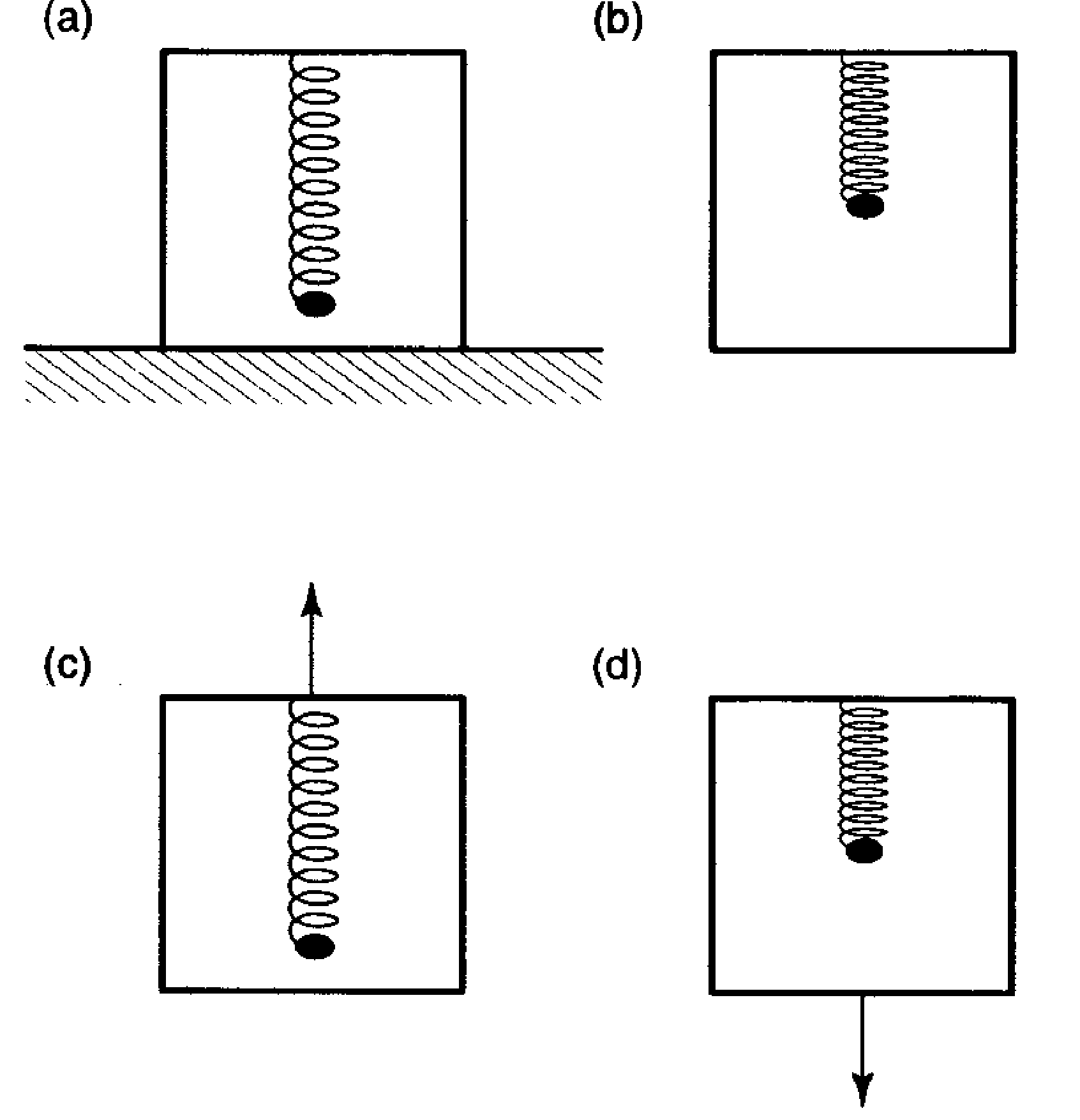}}
\caption{Thought-experiment illustrating
the equivalence principle. A weight is attached to a spring, which
is attached to the ceiling of a lift. In (a) the lift is
stationary, but a gravitational force acts downwards; the spring
is extended by the weight. In (b) the lift is in deep space, away
from any sources of gravity, and is not accelerated; the spring
does not extend. In (c) there is no gravitational field, but the
lift is accelerated upwards by a rocket; the spring is extended.
The acceleration in (c) produces the same effect as the
gravitational force in (a). In (d) the lift is freely-falling in a
gravitational field, accelerating downwards so no gravity is felt
inside; the spring does not extend because in this case the weight
is weightless and the situation is equivalent to (b).}
\end{figure}

What happens to the spring? The answer is that the acceleration
makes the weight move in the reverse direction relative to the
lift, thus extending the spring towards the floor. (This is like
what happens when a car suddenly accelerates - the passenger's
head is flung backwards.) But this is just like what happened when
there was a gravitational field pulling the spring down. If the
lift carried on accelerating, the spring would remain extended,
just as if it were not accelerating but placed in a gravitational
field. Einstein's idea was that these situations do not merely
appear similar: they are completely indistinguishable. Any
experiment performed in an accelerated lift in space would give
exactly the same results as one performed in a lift upon which
gravity is acting. To complete the picture, now consider a lift
placed inside a region where gravity is acting, but which is
allowed to fall freely in the gravitational field. Everything
inside becomes weightless, and the spring is not extended. This is
equivalent to the situation in which the lift is at rest and where
no gravitational forces are acting. A freely-falling observer has
every reason to consider himself to be in a state of inertial
motion.

Einstein now knew how he should construct the general theory of
relativity. But it would take him another ten years to produce the
theory in its final form \cite{Eins15}. What he had to find
was a set of laws that could deal with any form of accelerated
motion and any form of gravitational effect. To do this he had to
learn about sophisticated mathematical techniques, such as tensor
analysis and Riemannian geometry, and to invent a formalism that
was truly general enough to describe all possible states of
motion. He got there in 1915, and his theory is embodied in the
expression
\begin{equation}
G_{\mu \nu}=\frac{8\pi G}{c^4} T_{\mu \nu}\,\,\,;
\end{equation}
the entities $G$ and $T$ are tensors defined with respect to
four-dimensional coordinates $x_\mu$. The left-hand-side consists of 
the Einstein tensor $G$ which describes the curvature of space and the
tensor $T$ is the energy-momentum tensor which describes the
motion and properties of matter. Understanding the technicalities
of the general theory of relativity is a truly daunting task, and
calculating anything useful using the full theory is beyond all
but the most dedicated specialists. While the application of
Newton's theory of gravity requires one equation to be solved,
Einstein's theory (4) represents ten independent
equations which are all non-linear. Because of the equivalence
between mass and energy embodied in special relativity through
Equation (3), all forms of energy gravitate. The gravitational
field produced by a body is itself a form of energy, and it also
therefore gravitates. This non-linearity leads to unmanageable
mathematical complexity when it comes to solving the equations.
But the crucial aspect of this theory is that it relates the
properties and distribution of matter to the curvature of space.
This is what the 1919 expeditions were intended to test.

\section{The Bending of Light: From Principles to Principe}

\subsection{Curvature and the Equivalence Principle}
 The idea that space could be warped is so difficult to grasp that even
physicists don't really try to visualise such a thing. Our
understanding of the geometrical properties of the natural world
is based on the achievements of generations of Greek
mathematicians, notably the formalised system of Euclid -
Pythagoras' theorem, parallel lines never meeting, the sum of the
angles of a triangle adding up to 180 degrees, and so on. All of
these rules find their place in the canon of Euclidean geometry.
But these laws and theorems are not just abstract mathematics. We
know from everyday experience that they describe the properties of
the physical world extremely well. Euclid's laws are used every
day by architects, surveyors, designers and cartographers -
anyone, in fact, who is concerned with the properties of shape,
and the positioning of objects in space. Geometry is real.

It seems self-evident, therefore, that these properties of space
that we have grown up with should apply beyond the confines of our
buildings and the lands we survey. They should apply to the
Universe as a whole. Euclid's laws must be built into the fabric
of the world. Or must they? Although Euclid's laws are
mathematically elegant and logically compelling, they are not the
only set of rules that can be used to build a system of geometry.
Mathematicians of the 19th century, such as Gauss and Riemann,
realised that Euclid's laws represent only a special case of
geometry wherein space is flat. Different systems can be
constructed in which these laws are violated.

Consider, for example, a triangle drawn on a flat sheet of paper.
Euclid's theorems apply here, so the sum of the internal angles of
this triangle must be 180 degrees (equivalent to two
right-angles). But now think about what happens if you draw a
triangle on a sphere instead. It is quite possible to draw a
triangle on a sphere that has three right angles in it. For
example, draw one point at the 'north pole' and two on the
'equator' separated by one quarter of the circumference. These
three points form a triangle with three right angles that violates
Euclidean geometry.

Thinking this way works fine for two-dimensional geometry, but our
world has three dimensions of space. Imagining a three-dimensional
curved surface is much more difficult. But in any case it is
probably a mistake to think of 'space' at all. After all, one
can't measure space. What one can measure are distances between
objects located in space using rulers or, more realistically in an
astronomical context, light beams. Thinking of space as a flat or
curved piece of paper encourages one to think of it as a tangible
thing in itself, rather than simply as where the tangible things
are not. Einstein always tried to avoid dealing with entities such
as 'space' whose category of existence was unclear. He preferred
to reason instead about what an observer could actually measure
with a given experiment.

\begin{figure}
\centering
\resizebox*{8cm}{!}{\includegraphics{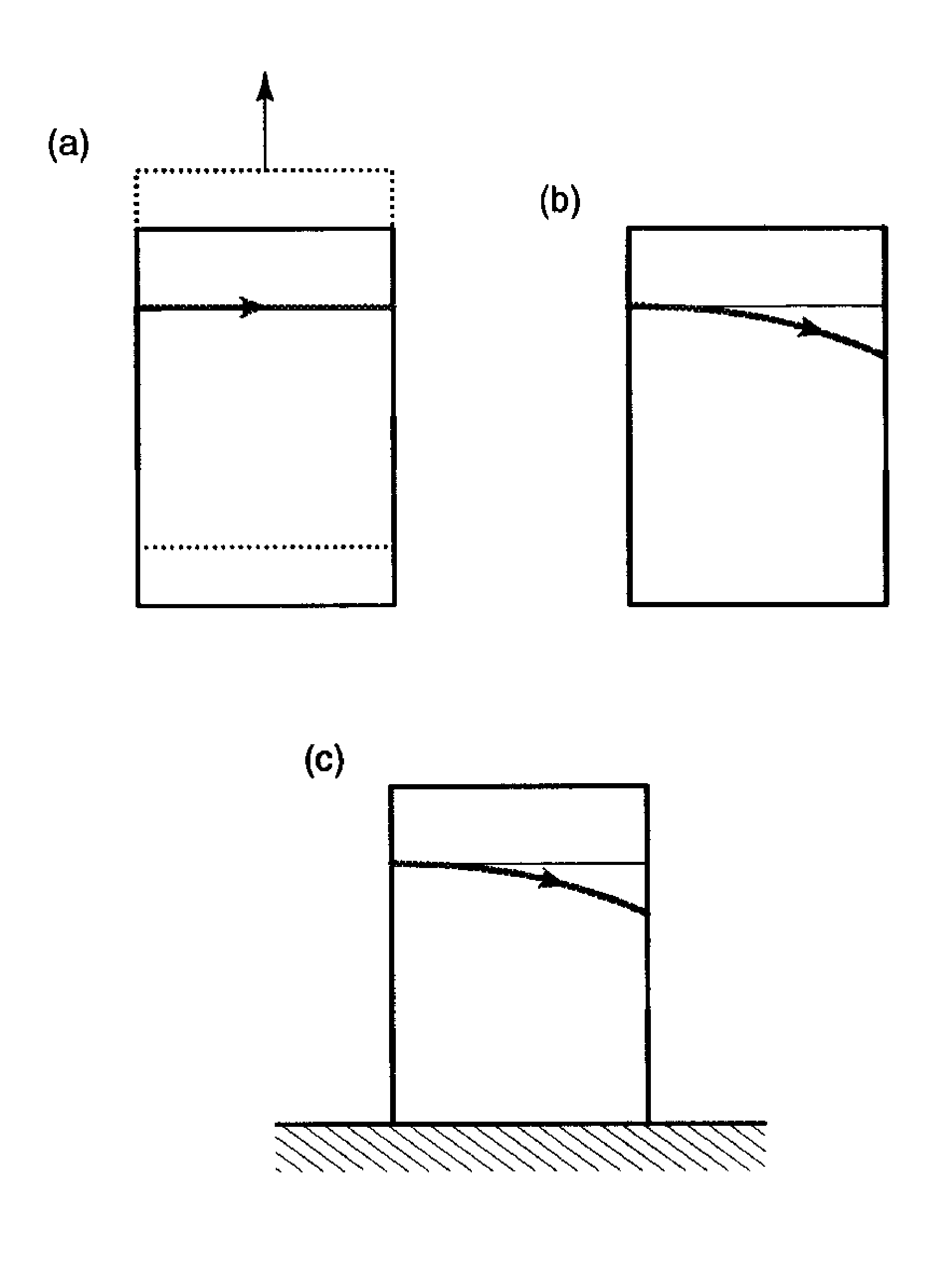}}
 \caption{The bending of light. In (a), our lift is
accelerating upwards, as in Figure 1(c). Viewed from outside, a
laser beam follows a straight line. In (b), viewed inside the
lift, the light beam appears to curve downwards. The effect in a
stationary lift situated in a gravitational field is the same, as
we see in (c).}
\end{figure}

Following this lead, we can ask what kind of path light rays
follow according to the general theory of relativity. In Euclidean
geometry, light travels on straight lines. We can take the
straightness of light paths to mean essentially the same thing as
the flatness of space. In special relativity, light also travels
on straight lines, so space is flat in this view of the world too.
But remember that the general theory applies to accelerated
motion, or motion in the presence of gravitational effects. What
happens to light in this case? Let us go back to the thought
experiment involving the lift. Instead of a spring with a weight
on the end, the lift is now equipped with a laser beam that shines
from side to side. The lift is in deep space, far from any sources
of gravity. If the lift is stationary, or moving with constant
velocity, then the light beam hits the side of the lift exactly
opposite to the laser device that produces it. This is the
prediction of the special theory of relativity. But now imagine
the lift has a rocket which switches on and accelerates it
upwards. An observer outside the lift who is at rest sees the lift
accelerate away, but if he could see the laser beam from outside
it would still be straight. He is not accelerating, so the special
theory applies to what he sees. On the other hand, a physicist
inside the lift notices something strange. In the short time it
takes light to travel across, the lift's state of motion has
changed (it has accelerated). This means that the point at which
the laser beam hits the other wall is slightly below the starting
point on the other side. What has happened is that the
acceleration has 'bent' the light ray downwards.

Now remember the case of the spring and the equivalence principle.
What happens when there is no acceleration but there is a
gravitational field, is exactly the same as in an accelerated
lift. Consider now a lift standing on the Earth's surface. The
light ray must do exactly the same thing as in the accelerating
lift: it bends downward. The conclusion we are led to is that
gravity bends light. And if light paths are not straight but bent,
then space is not flat but curved.

\subsection{Newton and Soldner}
The story so far gives the impression that nobody before Einstein
considered the possibility that light could be bent. In fact, this
is not the case. It had been reasoned before, by none other than
Isaac Newton himself, that light might be bent by a massive
gravitating object. In a rhetorical question posed in his Opticks,
Newton wrote:
\begin{quotation}
``Do not Bodies act upon Light at a distance, and by their action
bend its Rays; and is not this action . . . strongest at the least
distance?''
\end{quotation}
In other words, he was arguing that light rays themselves should
feel the force of gravity according to the inverse-square law. As
far as we know, however, he never attempted to apply this idea to
anything that might be observed. Newton's query was addressed in
1801 by Johann Georg von Soldner \cite{Soldner}. His work was motivated by the
desire to know whether the bending of light rays might require
certain astronomical observations to be adjusted. He tackled the
problem using Newton's corpuscular theory of light, in which light
rays consist of a stream of tiny particles. It is clear that if
light does behave in this way, then the mass of each particle must
be very small. Soldner was able to use Newton's theory of gravity
to solve an example of  a ballistic scattering problem.

\begin{figure}
\centering
\resizebox*{10cm}{!}{\includegraphics{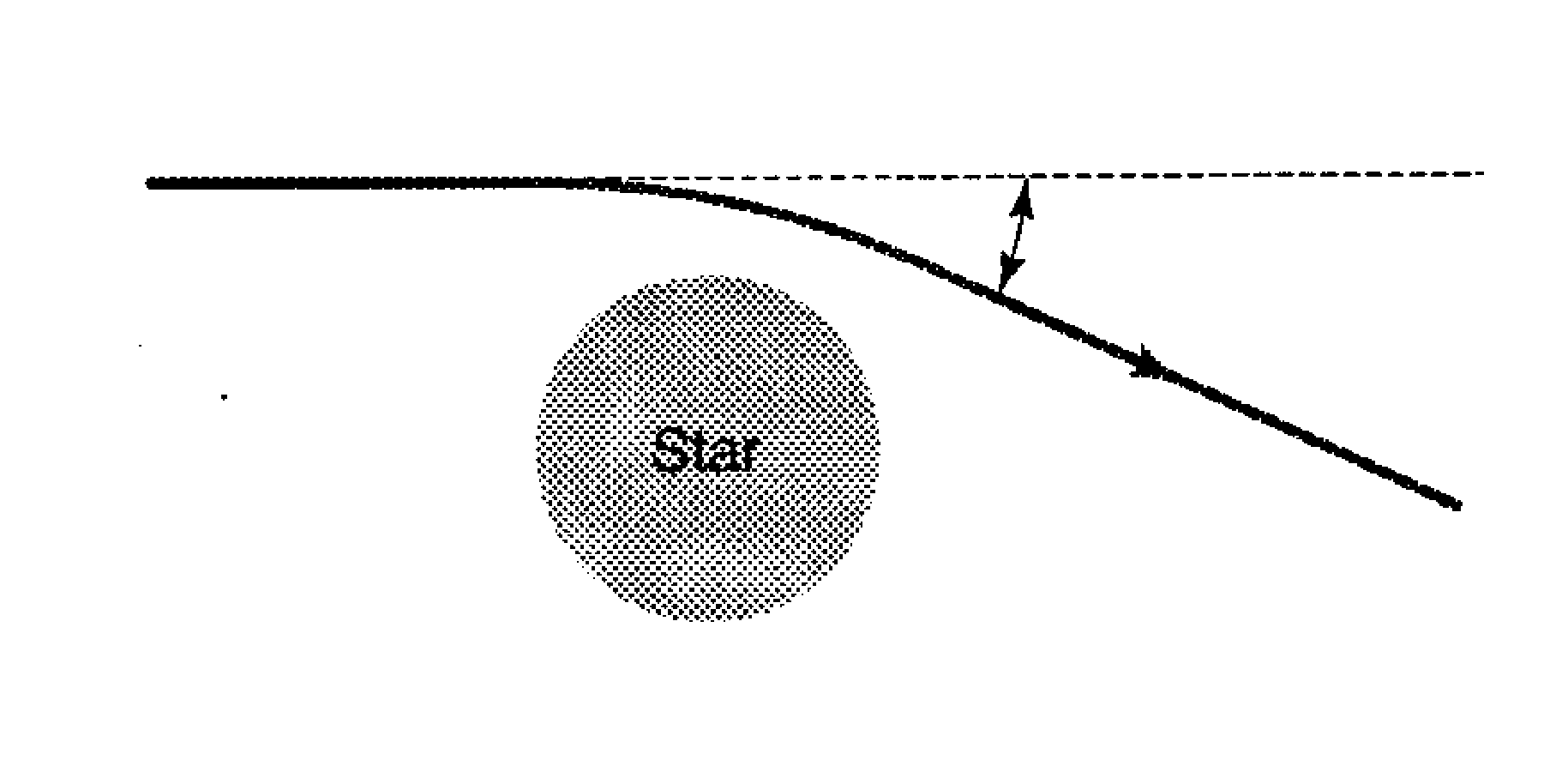}}
\caption{The ballistic scattering problem. A
small body passing close to a massive one is deflected through a
small angle on its way.}
\end{figure}

A small particle moving past a large gravitating object feels a
force from the object that is directed towards the centre of the
large object. If the particle is moving fast, so that the
encounter does not last very long, and the mass of the particle is
much less than the mass of the scattering body, what happens is
that the particle merely receives a sideways kick which slightly
alters the direction of its motion. The size of the kick, and the
consequent scattering angle, is quite easy to calculate because
the situation allows one to ignore the motion of the scatterer.
Although the two bodies exert equal and opposite forces on each
other, according to Newton's third law, the fact that the
scatterer has a much larger mass than the 'scatteree' means that
the former's acceleration is very much lower. This kind of
scattering effect is exploited by interplanetary probes, which can
change course without firing booster rockets by using the
gravitational 'slingshot' supplied by the Sun or larger planets.
When the deflection is small, the angle of deflection predicted by
Newtonian arguments, $\theta_N$ turns out to be
\begin{equation}
\theta_N=\frac{2GM}{rc^2},
\end{equation}
where $r$ is the `impact parameter', i.e. the closest distance to the scattering object
the incoming object would reach if it carried on along its initial trajectory.

Unfortunately, this calculation has a number of problems
associated with it. Chief amongst them is the small matter that
light does not actually possess mass at all. Although Newton had
hit the target with the idea that light consists of a stream of
particles, these photons, as they are now called, are known to be
massless. Newton's theory simply cannot be applied to massless
particles: they feel no gravitational force (because the force
depends on their mass) and they have no inertia, so you could argue that what
photons do in a Newtonian world is really anyone's guess. Nevertheless, the
Soldner result is usually called the Newtonian prediction, for
want of a better name.

\begin{figure}
\centering
\resizebox*{10cm}{!}{\includegraphics{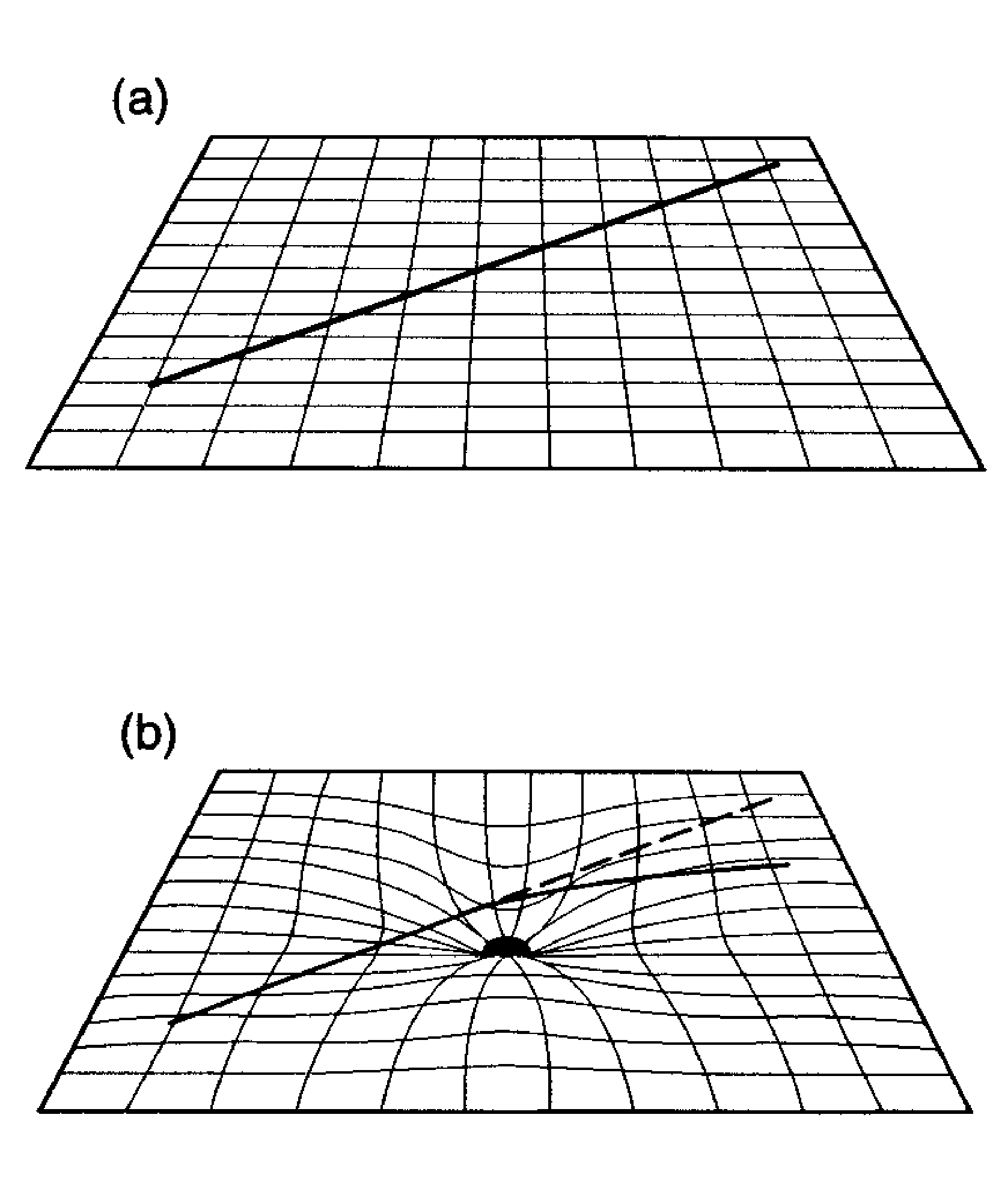}}
 \caption{Curved space and the bending of light. In
this illustration, space is represented as a two-dimensional
surface. In the absence of any gravitating bodies, light travels
in a straight line like a ball rolling on a smooth, flat table-top
(a). When a massive body is placed in the way, space becomes
curved: the closer you get to the body, the more curved it is (b).
The effect on light is as if the ball were rolling across a
table-top with a dip in the middle: it is deflected away from a
straight line.}
\end{figure}

Apparently unaware of Soldner's calculation, in 1907 Einstein began to think
about the possible bending of light. By this stage, he had already
arrived at the equivalence principle, but it was to be another
eight years before the general theory of relativity was completed.
He realised that the equivalence principle in itself required
light to be bent by gravitating bodies. But he assumed that the
effect was too small ever to be observed in practice, so he
shelved the calculation. In 1911, still before the general theory
was ready, he returned to the problem \cite{Eins11}. What he did in this
calculation was essentially to repeat the argument based on
Newtonian theory, but incorporating Equation (3). Although photons
don't have mass, they certainly have energy, and Einstein's theory
says that even pure energy has to behave in some ways like mass.
Using this argument, and spurred on by the realisation that the
light deflection he was thinking about might after all be
measurable, he calculated the bending of light from background
stars by the Sun.

For light just grazing the Sun's surface, i.e.for an impact parameter equal to
the radius of the Sun $r=R_\odot$ and
$M=M_\odot$, the mass of the Sun, Equation (5) yields a deflection of 0.87 seconds of
arc\footnote{For reference, one second of arc is roughly the angle subtended by the
width of a human hair at a distance of 10 metres}. This answer is precisely the same as the
Newtonian value obtained more than a century earlier by Soldner.

\begin{figure}
\centering
\resizebox*{10cm}{!}{\includegraphics{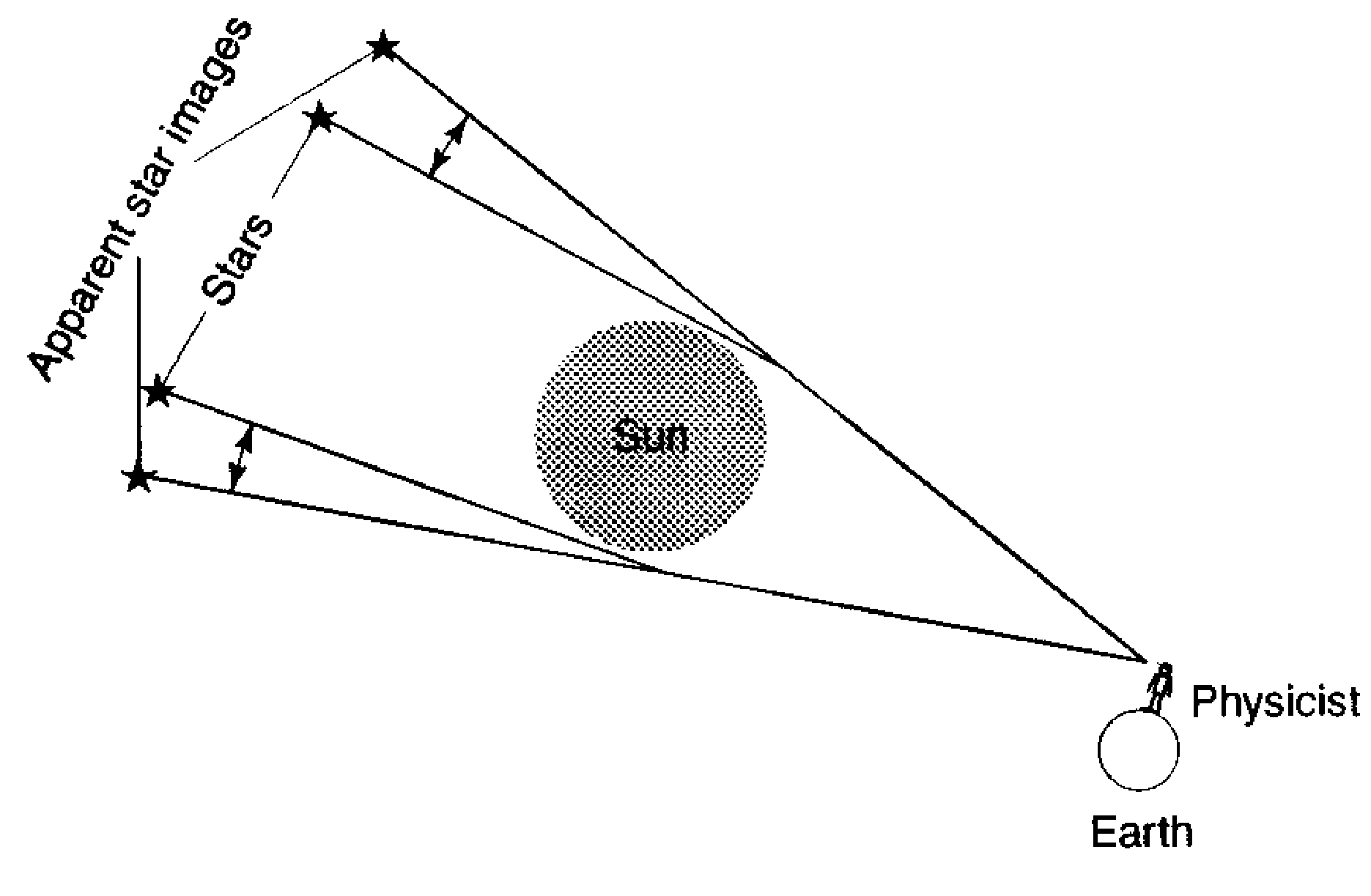}}
\caption{The bending of light by the Sun.
Light from background stars follows paths like those shown in
this Figure. The result is that the stars are seen in slightly
different positions in the sky when the Sun is in front of them,
compared to their positions when the Sun is elsewhere.}
\end{figure}

The predicted deflection is tiny, but according to the astronomers
Einstein consulted, it could just about be measured. Stars
appearing close to the Sun would appear to be in slightly
different positions in the sky than they would be when the Sun was
in another part of the sky. It was hoped that this kind of
observation could be used to test Einstein's theory. The only
problem was that the Sun would have to be edited out of the
picture, otherwise stars would not be visible close to it at all.
In order to get around this problem, the measurement would have to
be made at a very special time and place: during a total eclipse
of the Sun.

\subsection{The Relativistic Correction}

But this isn't quite where we take up the story of the famous
eclipse expeditions of 1919. There is a twist in the tale. In
1915, with the full general theory of relativity in hand, Einstein
returned to the light-bending problem. And he soon realised that
in 1911 he had made a mistake. The correct answer was not the same
as the Newtonian result, but twice as large.

What had happened was that Einstein had neglected to include all
effects of curved space in the earlier calculation. The origin of
the factor two is quite straightforward when one looks at how a
Newtonian gravitational potential distorts the metric of
space-time. In flat space (which holds for special relativity), and using
spherical coordinates, 
the infinitesimal interval in four dimensional space-time $ds$ is determined
by the relationship
\begin{equation}
ds^2=c^2dt^2-dr^2-r^2d\Omega^2,
\end{equation}
where $dt$ is the time interval, $dr$ is the radial distance and $d\Omega$
represents the element of solid angle. This is just a relativistic version of
Pythagoras' Theorem written in strange coordinates.
Light rays follow paths in space-time defined by $ds^2=0$ which
are straight lines in the absence of gravity, which is the case
in special relativity. In general relativity, the
theory is that light rays are no longer straight. In fact, around
a spherical distribution of mass $M$ the metric changes so that,
in the weak field limit, it becomes
\begin{equation}
ds^2=\left(1-\frac{2GM}{rc^2}\right)c^2 dt^2 -
\left(1-\frac{2GM}{rc^2}\right)^{-1}dr^2-r^2d\Omega^2.
\end{equation}
Since the corrections of order $GM/rc^2$ are small, as they are in this case,
one can solve the equation $ds^2=0$ by expanding each bracket in a power series
and keeping only the first term. In the weak field limit, the angular deflection predicted by Einstein's
equations is
\begin{equation}
\theta_E=\frac{4GM}{rc^2},
\end{equation}
which yields 1.74 arc seconds for $M=M_\odot$ and $r=R_\odot$,
compared to the 0.87 arc seconds obtained using Newtonian theory.
Not only is this easier to measure, being larger,  but it also
offers the possibility of a definitive test of the theory since it
differs from the Newtonian value.

Einstein's original calculation \cite{Eins11}
had included only the first bracket, corresponding
to the time component of the metric. The second, which arises from the space curvature, precisely doubles the net
deflection, at least in the limit of weak gravitational fields. It is worth noting that this is a particular
property of Einstein's theory: it is possible to construct theories of gravity in which the effects
on time and space are not equal in magnitude. For a fuller discussion of this and
other aspects of the light-bending problem, see \cite{Long}.

In 1912, an Argentinian expedition had been sent to Brazil to
observe a total eclipse. Light-bending measurements were on the
agenda, but bad weather prevented them making any observations. In
1914, a German expedition, organised by Erwin Freundlich and
funded by Krupp, the arms manufacturer, was sent to the Crimea to
observe the eclipse due on 21 August. But when the First World War
broke out, the party was warned off. Most returned home, but
others were detained in Russia. No results were obtained. The war
made further European expeditions impossible. \footnote{One wonders how
Einstein would have been treated by history if either of the 1912
or 1914 expeditions had been successful! Until 1915, his
reputation was riding on the incorrect value of 0.87 arc seconds.
As it turned out, the 1919 British expeditions to Sobral and
Principe were to prove his later calculation to be right.}

\section{Eddington and the Expeditions}
The story of the 1919 eclipse observations revolves around an astronomer by
the name of Arthur Stanley Eddington. His life and work is
described by Douglas \cite{Doug57}  and Chandrasekhar \cite{Chandra}.
Eddington was born in Cumbria in 1882, but moved with his mother to Somerset
in 1884 when his father died. He was brought up as a devout
Quaker, a fact that plays an important role in the story of the
eclipse expedition. In 1912, aged only 30, he became the Plumian
Professor of Astronomy and Experimental Philosophy at the
University of Cambridge, the most prestigious astronomy chair in
Britain, and two years later he became director of the Cambridge
observatories. Eddington had led an expedition to Brazil in 1912
to observe an eclipse, so his credentials made him an ideal
candidate to measure the predicted bending of light.

Eddington was in England when Einstein presented the general
theory of relativity to the Prussian Academy of Sciences in 1915 \cite{Eins15}.
Since Britain and Germany were at war at that time, there was no
direct communication of scientific results between the two
countries. But Eddington was fortunate in his friendship with the
astronomer  Willem De Sitter, later to become one of the founders
of modern cosmology, and who was in neutral Holland at the time.
De Sitter received copies of Einstein's papers, and wasted no time
in passing them onto Eddington in 1916. Eddington was impressed by
the beauty of Einstein's work, and immediately began to promote
it. In a report to the Royal Astronomical Society in early 1917,
he particularly stressed the importance of testing the theory
using measurements of light bending. A few weeks later, the
Astronomer Royal, Sir Frank Watson Dyson, realised that the
eclipse of 29 May 1919 was especially propitious for this task \cite{Dyson17}.
Although the path of totality ran across the Atlantic ocean from
Brazil to West Africa, the position of the Sun at the time would
be right in front of a prominent grouping of stars known as the
Hyades. When totality occurred, the sky behind the Sun would be
glittering with bright stars whose positions could be measured.
Moreover, at over seven minutes, the duration of totality was long enough
to enable the measurements to repeated several times at each location to check
for consistency.

Dyson began immediately to investigate possible observing sites.
It was decided to send not one, but two expeditions. One, led by
Eddington, was to travel to the island of Principe off the coast
of Spanish Guinea in West Africa, and the other, led by Andrew
Crommelin (an astronomer at the Royal Greenwich Observatory),
would travel to Sobral in northern Brazil. An application was made
to the Government Grant Committee to fund the expeditions, \pounds
100 for instruments and \pounds 1000 for travel and other costs.
Preparations began, but immediately ran into problems.

Although Britain and Germany had been at war since 1914,
conscription into the armed forces was not introduced in England
until 1916. At the age of 34, Eddington was eligible for the
draft in 1916 but, as a Quaker, he let it be known that he would refuse to
serve. The climate of public opinion was heavily against
conscientious objectors. Eddington might well have been sent with
other Quaker friends to a detention camp and spent the rest of the
war peeling potatoes. Dyson, and other prominent Cambridge
academics, went to the Home Office to argue that it could not be
in the nation's interest to have such an eminent scientist killed
in the trenches of the Somme. After much political wrangling, a
compromise was reached. Eddington's draft was postponed, but only
on condition that if the war ended by 29 May 1919, he must lead
the expedition to Principe.

Even with this hurdle out of the way, significant problems
remained. The expeditions would have to take specialised
telescopes and photographic equipment. But the required
instrument-makers had either been conscripted or were engaged in
war work. Virtually nothing could be done until the armistice was
signed in November 1918. Preparations were hectic, for the
expeditions would have to set sail in February 1919 in order to
arrive and set up camp in good time. Moreover, reference plates
would have to be made. The experiment required two sets of
photographs of the appropriate stars. One, of course, would be
made during the eclipse, but the other set (the reference plates)
had to be made when the Sun was nowhere near that part of the sky.
In order to correct for possible systematic effects the reference
plates should ideally be taken at the same site and at the same
elevation in the sky: this would mean waiting at the observation
site until the stars behind the Sun during the eclipse would be at
the same position in the sky before dawn. This was not too much of
a problem at Sobral, where the eclipse occurred in the morning but
at Principe, Eddington would have had to wait for several months to
take his reference plates. He was therefore forced to rely on comparison plates
taken at Oxford before his departure for Principe.

\begin{figure}
\centering
\resizebox*{10cm}{!}{\includegraphics{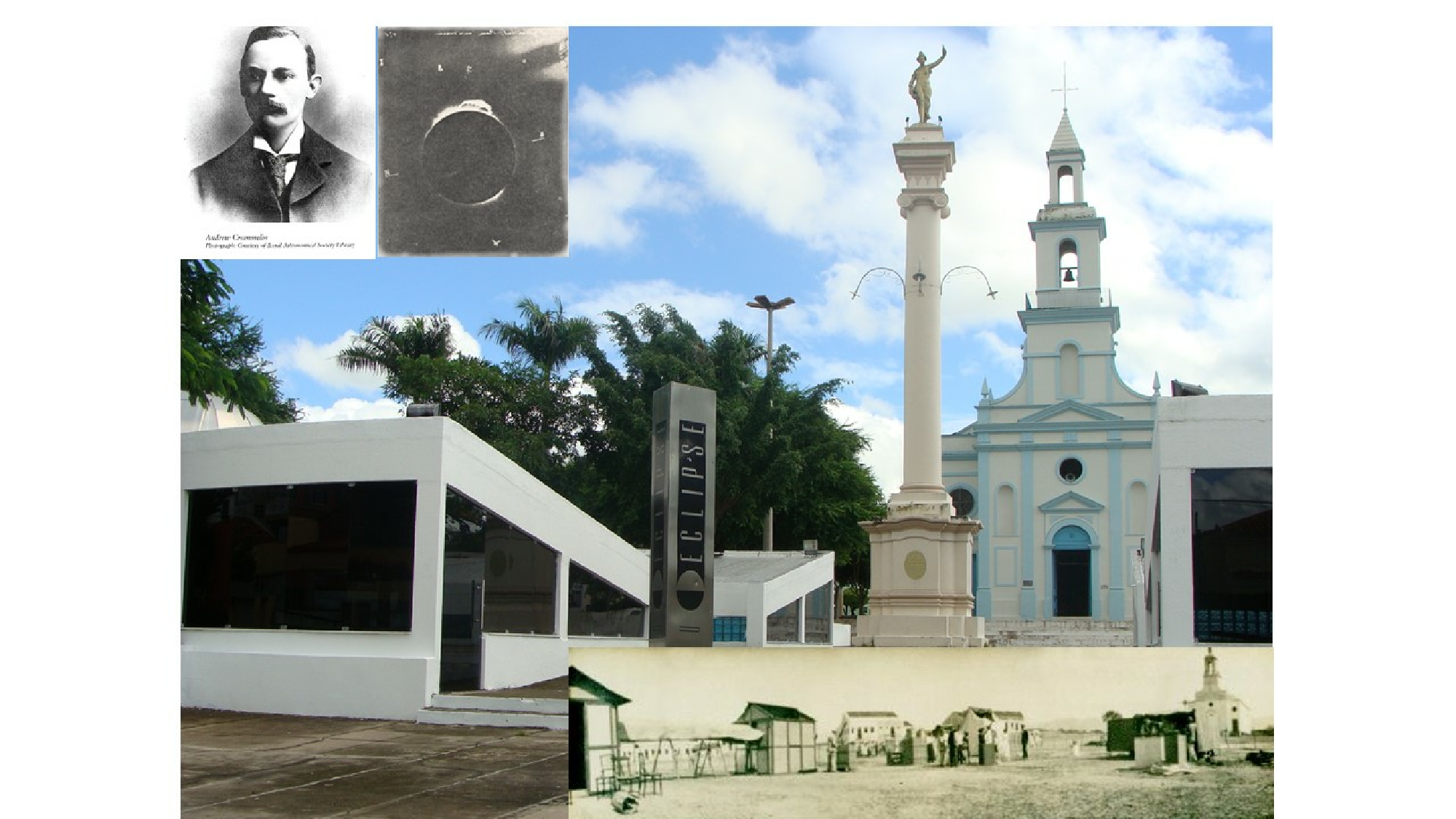}}
\caption{The modern town of Sobral in the state of Ceara, Brazil,contains
a museum dedicated to the eclipse placed close to the site
where the 1919 observers placed their experiments (see 
same church in lower right panel). The British campaign
was led by Andrew Crommelin from the Royal Greenwich
Observatory shown top left with one of the original Sobral eclipse plates.}
\end{figure}

\begin{figure}
\centering
\resizebox*{10cm}{!}{\includegraphics{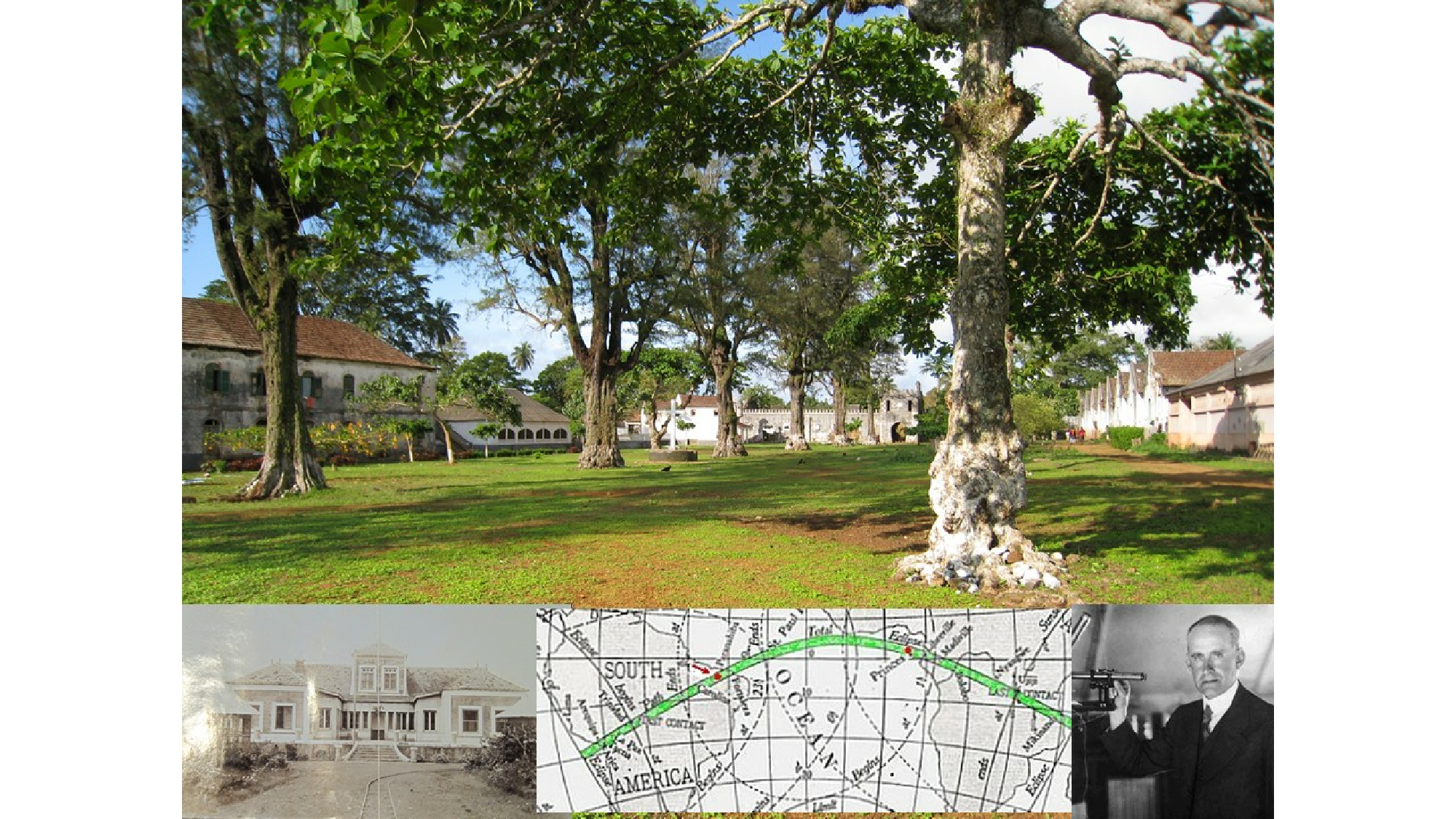}}
\caption{The cocoa plantation Roca Sundy on the island of Principe where Eddington
and his assistant Cottingham undertook eclipse observations. At
lower left is the house where they stayed which now stands
close to various plaques commemorating the expedition.
Lower centre shows the path of totality which reached
Principe in the early afternoon. Eddington (lower right) was 36 at
the time.}
\end{figure}

In the end the expeditions set off on time (in February 1919) and
back home the astronomical community, particularly in Britain,
chewed its collective nails. There were several possible outcomes.
They might fail to measure anything, due to bad weather or some
other mishap. They might measure no deflection at all, which would
contradict all the theoretical ideas of the time. They might find
the Newtonian value, which would humiliate Einstein. Or they might
vindicate him, by measuring the crucial factor of two. Which would
it be?

The June 1919 issue of Observatory magazine, which carries news of
Royal Astronomical Society meetings and certain other matters,
contains a Stop-Press item. Two telegrams had arrived. One was
from Crommelin in Sobral: 'ECLIPSE SPLENDID'. The other, from
Eddington, was disappointing: 'THROUGH CLOUD. HOPEFUL'. The
expeditions returned and began to analyse their data. The
community waited.

\begin{figure}
\centering
\resizebox*{10cm}{!}{\includegraphics{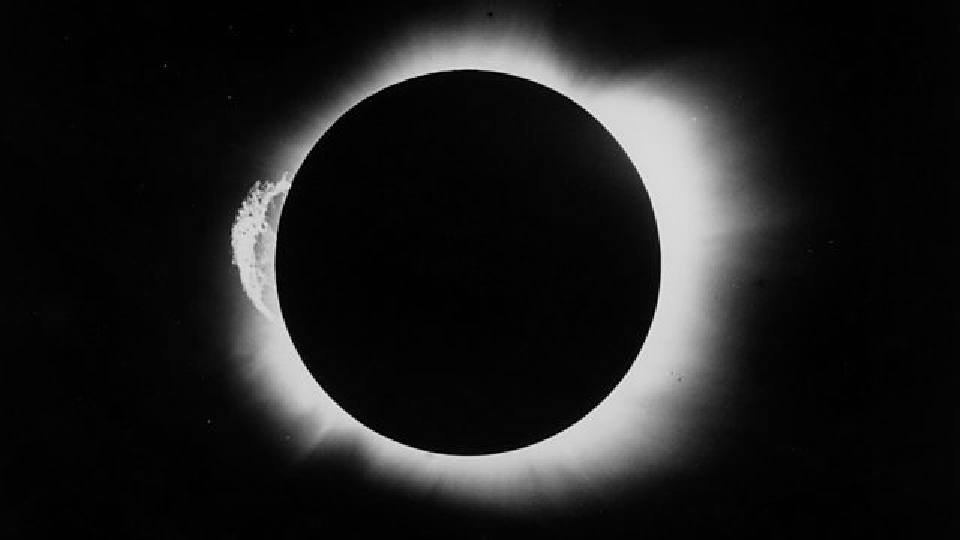}}
\caption{A photograph of the 1919 eclipse, taken at Principe, showing a spectacular prominence.}
\end{figure}

\subsection{Measurement and Error}

The full details of both expeditions can be read in the account
published in {\em Philosophical Transactions of the Royal Society}
\cite{DED20}. The main items of experimental equipment were
two astrographic object glasses of about 10 inches in diameter,
one from Oxford and one from the Royal Greenwich Observatory. 
These lenses, made by Howard Grubb's works in Dublin for
the {\em Carte Du Ciel} international photographic star catalogue project,
were specially designed to measure star positions over
a relatively large patch of the sky and were therefore ideal for
the kind of experiment being done during the eclipse. The
objective lenses  were removed from the observatories in which they were
usually housed (in Oxford and Greenwich) and steel tubes were built to form temporary
telescopes for the expeditions. Almost as an afterthought, it was
decided to take a much smaller objective lens, 4 inches in
diameter, along with a coelostat belong to the Royal Irish Academy,
to the Sobral site as a kind of backup. Along with these very important Irish
contributions to the equipment, it is worth noting that Andrew Crommelin was
himself born in County Antrim.

The expeditions also took two large coelostats, mirrors used
especially for solar observations. The reason for the mirrors was
that no mechanical devices were available to drive the steel tubes
containing the object glasses to compensate for the rotation of
the Earth. The tubes had to be as long as the focal length of the
lens, which in this case was about 3.5 metres, so they were
difficult to move once set up. If a telescope is not moved by such
a driver during the taking of a photograph, the stars move on the
sky during the exposure and the images turn into streaks. In the
eclipse experiment, the trick used was to keep the telescope,
still but to have it pointing downwards towards the coelostat
which reflects the light into the telescope lens. The mirror is
much small (about 16 inches across) and a relatively small
clockwork device can be used to move it to correct for the Earth's
rotation instead of moving the whole telescope.

It is clear that both expeditions had encountered numerous
technical problems. The day of the eclipse arrived at Principe
with heavy cloud and rain. Eddington was almost washed out, but
near totality the Sun began to appear dimly through cloud and some
photographic images could be taken. Most of these were
unsuccessful, but the Principe mission did manage to return with
two useable photographic plates. Sobral had better weather but
Crommelin and  his team had set the focus of their main telescope
overnight before the eclipse, when there were plenty of bright stars around to check its
optical performance. When the day of the
eclipse dawned and the temperature began to rise, he  watched
with growing alarm as both the steel tube and the coelostat mirror
began to expand with the heat. As a result, most of the main
Sobral plates were badly blurred and distorted by astigmatism; there
was also some confusion caused by the solar corona for stars
very close to the Sun's disk.

On the other hand, the little 4-inch telescope taken as a backup, which was mounted in a wooden tube,
performed very well and the plates obtained with it were to prove the most
convincing in the final analysis.

There were other problems too. The light deflection expected was
quite small: less than two seconds of arc. But other things could
cause a shifting of the stars' position on a photographic plate.
For one thing, photographic plates can expand and contract with
changes in temperature. The emulsion used might not be
particularly uniform. The eclipse plates might have been exposed
under different conditions from the reference plates, and so on.
The Sobral team in particular realised that, having risen during
the morning, the temperature fell noticeably during totality, with
the probable result that the photographic plates would shrink. The
refractive properties of the atmosphere also change during an
eclipse, leading to a false distortion of the images. And perhaps
most critically of all, Eddington's expedition was hampered by bad
luck even after the eclipse. Because of an imminent strike of the
local steamship operators, his team was in danger of being
completely stranded. He was therefore forced to leave early,
before taking any reference plates of the same region of the sky
with the same equipment. Instead he relied on one check plate made
at Principe and others taken previously at Oxford. These were
better than nothing, but made it impossible to check fully for
systematic errors and laid his results open to considerable
criticism.

\subsection{A Bit of Data Analysis}

It is worth spelling out in a little more detail how the plate comparison
is done. Suppose one can measure for a given star the difference between
its position on the comparison plate (with no effect of the Sun's gravity) and an eclipse plate
(where the Sun's gravity deflects light). This difference can be in any direction on the plate
so must be expressed in terms of two components (say Right Ascension and Declination):
\begin{eqnarray}
  D_x & = & ax + by+c+ \alpha E_x(x,y)\nonumber\\
  D_y & = & dx +ey+f+\alpha E_y(x,y).
\end{eqnarray}
Here $x$ and $y$ are the actual coordinates of a star on the plate. The right hand side $D_x$ and $D_y$ are the measured deflections.
The terms $E_x(x,y)$ and $E_y(x,y)$ are the two componenets of the gravitational deflection expected at the star's position; recall that
the deflection is radially outwards from centre of the Sun. The constant $\alpha$  would be $\alpha=0.87$ for
`Newtonian' deflection and $\alpha=1.74$ for Einstein's theory. The direction of the deflection being the same in both theories, $\alpha$ is the key
 parameter to be determine from the measurements.

 However, gravitational deflection is not the only thing that could cause slight differences in the positions of stars on two
 photographic plates. The other terms in the above equations are intended to model these. The constants $c$ and $f$ are offsets that could be caused by incorrect centring
 of the plates. The terms $a$ and $e$ represent systematic errors in position that are proportional to the position itself. These correspond to an error
 in scale or magnification. The terms $b$ and $d$ correspond to errors in $x$ proportional to $y$ (and vice versa); these would arise
 if the plates were not quite oriented in the same way. One can do one's best to reduce these sources of error but they will never be eliminated completely.
 In order to isolate the effect of the gravitational deflection the only way to proceed is to measure sufficient star positions to estimate
 each parameter, in other words to obtain an astrometric solution. Only when this has been done can one estimate $\alpha$.

Scientific observations are always subject to errors and
uncertainty. The level of this uncertainty in any
experimental result is usually communicated in the technical
literature by giving not just one number as the answer, but
attaching to it another number called the 'standard error', an
estimate of the range of possible errors that could influence the
result. If the light deflection measured was, say, 1 arc second,
then this measurement would be totally unreliable if the standard
error were as large as the measurement itself, 1 arc second. Such
a result would be presented as '$1 \pm 1$' arc second, and nobody
would believe it because the measured deflection could well be
produced entirely by instrumental errors. In fact, as a rule of
thumb, physicists never usually believe anything unless the
measured number is larger than two standard errors. The expedition
teams analysed their data, with Eddington playing the leading
role, cross-checked with the reference plates, checked and
double-checked their standard errors. Finally, they were ready.

\subsection{Results and Reaction}
A special joint meeting of the Royal Astronomical Society and the
Royal Society of London was convened on 6 November 1919. Dyson
presented the main results, and was followed by contributions from
Crommelin and Eddington. The results from Sobral, with
measurements of seven stars in good visibility, gave the
deflection as $1.98 \pm 0.18$ arc seconds. Principe was less
convincing. Only five stars were included, and the conditions
there led to a much larger error. Nevertheless, the value obtained
by Eddington was $1.62 \pm 0.45$. Both were within two standard
errors of the Einstein value of 1.74 and more than two standard
errors away from either zero or the Newtonian value of 0.87.

The reaction from scientists at this special meeting was
ambivalent. Some questioned the reliability of statistical
evidence from such a small number of stars. This skepticism seems
in retrospect to be entirely justified. Although the results from
Sobral were consistent with Einstein's prediction, Eddington had
been careful to remove from the analysis all measurements taken
with the main equipment, the astrographic telescope and used only
the results from the 4-inch. As I have explained, there were good
grounds for this because of problems with the focus of the larger
instrument. On the other hand, these plates yielded a value for
the deflection of about  0.93 seconds of arc, very close to the Newtonian
prediction. Some suspected Eddington of cooking the books by
leaving these measurements out. For a full discussion of the controversy,
see \cite{Kenn09}.

In any case, as it happens, the plates
produced by the Sobral astrographic were remeasured using
an automated plate-measuring device in 1979. This type of
device is less likely to be confused by distortion than the human eye. After the
remeasurement \cite{Harvey}, the results from the Sobral
astrographic were $\alpha=1.55 \pm 0.34$, which is consistent
with the Einstein value.

It is a great shame that the original photographs taken during the 1919
eclipse are not available to be measured and reanalysed. It seems the Principe
plates were thrown when Eddington died in 1944, possibly by his sister with whom he
lived at the Observatory House in Cambridge and who had to move out after his death.
The Sobral plates were definitely available in 1979 for the re-measurement
discussed above, but they also seem to have been lost - perhaps as a consequence
of the numerous reorganisations of the Royal Observatories over the past few decades.

Although the plates are not available, at least the measurements made in 1919
are tabulated fully in \cite{DED20} and it is not too difficult a task
to re-do the analysis. In fact I have set this task as an undergraduate exercise
for many years as a good example of the use of statistics in astrophysics. 

Opinion seems to have been divided among the audience at the Royal Society.
Ludwick Silberstein, admonished the audience: he pointed a finger
at the portrait of Newton that hangs in the meeting room, and
warned: 'We owe it to that great man to proceed very carefully in
modifying or retouching his Law of Gravitation.' On the other
hand, the eminent Professor J.J. Thomson, discoverer of the
electron and Chair of the meeting, was convinced, stating
\begin{quotation}
``This is the most important result obtained in connection with
the theory of gravitation since Newton's day.''
\end{quotation}
Einstein himself had no doubts. He had known about the results
from the English expeditions before the formal announcement in
November 1919. On 27 September, he had written an excited postcard
to his mother:
\begin{quotation}
``. . . joyous news today. H.A. Lorentz telegraphed that the
English expeditions have actually measured the deflection of
starlight from the Sun.''
\end{quotation}
He later down-played his excitement in a puckish remark about his
friend and colleague, the physicist Max Planck: \begin{quotation}
``He was one of the finest people I have ever known . . . but he
didn't really understand physics, [because] during the eclipse of
1919 he stayed up all night to see if it would confirm the bending
of light by the gravitational field. If he had really understood
[the general theory of relativity], he would have gone to bed the
way I did.''
\end{quotation}
In 1922, another eclipse, viewed this time from Australia, yielded
not a handful, but scores of measured position-shifts and much
more convincing statistical data \cite{CT23}).
Even so, the standard error
on these later measurements was of similar size, around 0.20 arc
seconds. Measurements of this kind using optical telescopes to
measure light deflection continued until the 1950s, but never
increased much in accuracy because of the fundamental problems in
observing stars through the Earth's atmosphere. More recently,
similar measurements have been made, not using optical light but
radio waves. These have the advantage that they are not scattered
by the atmosphere like optical light is. The light-bending
measurement for radio sources rather than stars can be made almost
at will, without having to wait for an eclipse. For example, every
a quasar passes behind the Sun it produces a
measurable deflection. These measurements confirm the Einstein
prediction and it is now accepted by the vast majority of
physicists that light is bent in the manner suggested by the
general theory of relativity; a quantitative summary of many
experimental tests can be found in \cite{Bert62}.

Moreover, other predictions of the general theory also seem to fit
with  observations:  the orbital spin of the binary pulsar
is a notable example because it led to the award of
the 1993 Nobel Prize. More recently we have now seen firm evidence
of the existence of gravitational waves, another key prediction of Einstein's theory,
also resulting in a Nobel Prize (in 2017). There might have been controversy in 1919, but a hundred years on, it seems
that general relativity is firmly established as a sound scientific theory.

\section{Discussion}

The eclipse expeditions of 1919 certainly led to the eventual
acceptance of Einstein's general theory of relativity in the
scientific community. This theory is now an important part of the
training of any physicist and is regarded as the best we have for
describing the various phenomena attributable to the action of
gravity. The events of 1919 also established Einstein, rightly, as
one of the century's greatest intellects. But it was to do much
more than that, propelling him from the rarefied world of
theoretical physics into the domain of popular culture. How did
this happen?

Einstein, and his theory of relativity, had appeared in newspapers
before 1919, mainly in the German-speaking world. He had himself
written an article for Die {\em Vossische Zeitung} in 1914. But he
had never experienced anything like the press reaction to the
announcements at the Royal Society meeting in 1919. Indeed, as
Abraham Pais notes in his superb biography of Einstein, the New
York Times index records no mention at all of Einstein until 9
November 1919. From then until his death in 1955, not a year
passed without a mention of Einstein's name.

Some of the initial attention also rubbed off on Eddington. He ran
a series of lectures in Cambridge on Einstein's theory. Hundreds
turned up and the lectures were packed. Eddington became one of
the foremost proponents of the new theory in England, and went on
to inspire a generation of astrophysicists in Cambridge and
beyond. But this was nothing compared to what happened to Albert
Einstein.

The London Times of 7 November 1919 carried a long article about
the Royal Society meeting, headlined 'REVOLUTION IN SCIENCE. NEW
THEORY OF THE UNIVERSE'. Two days later, the New York Times
appeared with the headline 'LIGHTS ALL ASKEW IN THE HEAVENS'. But
these splashes were not to be short-lived. Day after day, the
global media ran editorials and further features about Einstein
and his theory. The man himself was asked to write an article for
the London Times, an offer he accepted 'with joy and
gratefulness'. Gradually, the press reinforced the role of
Einstein as genius and hero, taking pains to position him on one
side of an enormous intellectual gulf separating him from the
common man. He emerged as a saintly, almost mythical character who
was afforded great respect by scientists and non-scientists alike.

As years passed his fame expanded further still, into parts of
popular culture that scientists had never occupied before.
Einstein was invited to appear in Variety at the London Palladium
(doing what one can only guess). He featured in popular songs,
movies and advertisments. Eventually, this attention wore him
down. Towards the end of his life he wrote to a friend:
\begin{quotation}
``Because of the peculiar popularity which I have acquired,
anything I do is likely to develop into a ridiculous comedy. This
means that I have to stay close to home and rarely leave
Princeton.''
\end{quotation}
No scientist working today would begrudge the fame that settled on
Einstein. His achievements were stunning, with all the hallmarks
of genius stamped upon them. But while his scientific
contributions were clearly a necessary part of his canonisation,
they are not sufficient to explain the unprecedented public
reaction.

One of the other factors that played a role in this process is
obvious when one looks at the other stories in the London Times of
7 November 1919. On the same page as the eclipse report, one finds
the following headlines: 'ARMISTICE AND TREATY TERMS'; 'GERMANS
SUMMONED TO PARIS'; 'RECONSTRUCTION PROGRESS'; and 'WAR CRIMES
AGAINST SERBIA'. To a world wearied by a terrible war, and still
suffering in its aftermath, this funny little man and his crazy
theories must have been a welcome distraction, even if his ideas
themselves went way over the heads of ordinary people. Here too
was token of a much-needed reconciliation between Britain and
Germany. In his Times article, Einstein stressed that science cuts
across mere national boundaries, hinting that if politicians
behaved more like scientists there would be no more pointless
destruction on the scale that Europe had just experienced.

Whatever the reason for Einstein's personal fame, there is no question
that the 1919 eclipse expeditions ushered in a new era of scientific progress.
The measurement of the bending of light by the Sun's gravity was an experiment
that changed the world, or at least our perception of it, overthrowing
Newton's concepts of absolute time and absolute space.
It truly was a revolution in science.

\section*{Acknowledgments}
I am grateful to Richard Ellis for the original photographs of the Sobral
and Principe locations shown in Figures 5 and 6 respectively. I also acknowledge
very helpful conversations on the topic of this paper with Robin Catchpole of the Royal Observatory
Greenwich and Tom Ray of Dublin Institute for Advanced Studies.
This piece is a greatly expanded version of \cite{PC2001}.

\end{document}